\documentclass[preprint,aps,floats]{revtex4}

\def\d{\mathrm{d}}

\begin{document}

\title{Self-consistent theory of compact ${\rm QED}_3$
with relativistic fermions}

\author{Matthew J. Case, Babak H. Seradjeh and Igor F. Herbut}

\affiliation{Department of Physics, Simon Fraser University\\
Burnaby, British Columbia, Canada V5A 1S6}

\begin{abstract}
We consider three dimensional quantum
electrodynamics (${\rm cQED}_3$) with massless relativistic
fermions coupled to a compact gauge field using a combined perturbative
variational approach. Coupling to matter renders the bare interaction between
magnetic monopoles logarithmic at large distances, suggesting 
the possibility of a confinement-deconfinement transition
of the Berezinskii-Kosterlitz-Thouless type in the theory. Our self-consistent
calculation suggests, however, that
screening effects always destabilise the confined phase, in
agreement with the previous renormalisation group study of the same
model.
\end{abstract}

\maketitle

\section{Introduction}

Compact U(1) gauge theories in three ($d=3$)
 dimensions have long been of interest in
high energy and condensed matter physics.  In particle
physics they serve as relatively simple models exhibiting non-perturbative
phenomena such as chiral symmetry breaking and confinement~\cite{Polyakov77,
Pisarski84,Appelquist86}, believed to be crucial
to our understanding of more realistic theories like
quantum chromodynamics.  In condensed matter physics the 
theories with compact U(1) gauge fields coupled to matter arise frequently
in descriptions of strongly correlated electron systems~\cite{Kim99}. In 
this case the three dimensional models are of direct significance to
condensed matter systems in two ($d=2$) spatial
dimensions and at zero temperature ($T=0$).

A crucial issue in all compact U(1) theories  is the 
confinement of `charge' due to the unbinding of magnetic monopoles, which 
are invariably introduced by the compact nature of the gauge field.  In a
pioneering work,  Polyakov~\cite{Polyakov77} showed that in pure
compact quantum electrodynamics  without matter in $d=3$ 
confinement is permanent for all values of the gauge
coupling.  The situation where the gauge field is coupled to
matter is more subtle, and a subject of current debate. 
It has been argued that coupling to relativistic
massless fermions transforms the usual
Coulombic interaction between monopoles into the much longer-ranged
logarithmic interaction at
large distances~\cite{Ioffe89, Rantner01, Wen02, Kleinert02}.
When applied to a single monopole-antimonopole pair, this 
would suggest that monopoles may  bind into dipoles,
in analogy with the celebrated Berezinskii~\cite{B}, Kosterlitz and
Thouless~\cite{KT} (BKT) transition in two dimensions. However, while
the effects of a finite density of monopoles on the BKT transition in
$d=2$ are well understood~\cite{KT, Kosterlitz74}, the
situation in $d=3$ appears less clear~\cite{Kosterlitz77}.
The difficulty lies in the fact that while the screening  in the dipole
phase in $d=2$ just amounts to renormalisation of the
dielectric constant, in $d=3$ it changes the {\it form} of the
interaction~\cite{Murthy91, Sachdev02, Igor03, Subir03}.
In a recent paper, two of us~\cite{Igor03} presented an
electrostatic argument and a renormalisation group calculation to
show that the interaction between distant monopoles in the presence
of other dipoles is screened back into the Coulomb potential. Together
with the generalisation to
the case of coupling to non-relativistic fermions~\cite{Subir03},
this strongly suggests that the putative deconfined phase 
in $d=3$ is always unstable. Compact U(1) theories in $d=3$, with or without
matter, would appear therefore generically to be permanently confining. 

In the present article we study the issue of confinement in 
cQED$_3$ using the variational treatment of the anomalous
sine-Gordon (ASG) theory, which
is dual to the original ${\rm cQED}_3$.  By working to the
second order in fugacity and  including the screening
effects we find that monopoles are
free at any effective temperature in the ASG theory
(i.~e. for  any number of fermion flavours in cQED$_3$).
This suggests that fermions are permanently in the
confined phase, and provides an additional support to the
renormalisation group results of Refs.~\cite{Igor03} and~\cite{Subir03}.

We introduce the ${\rm cQED_3}$ theory and its dual sine-Gordon version
in Section~\ref{model}.  In
Section~\ref{sGBF}, we discuss the lowest order variational
calculation that neglects screening and point to its limitations.
We then propose a generalised self-consistent 
approach that includes higher orders in monopole fugacity and allows 
for the screening effects in Section~\ref{SCPA}. In Section~\ref{results}
we present the calculations to the second
order.  A summary of our results is given in Section~\ref{conc}.

\section{${\rm cQED_3}$ and the anomalous sine-Gordon theory}
\label{model}

We will be interested in the phases of ${\rm cQED}_3$, with
the gauge field coupled to massless relativistic fermions on a lattice:
 \begin{equation}
 \label{Slatt}
   S[\chi, a] = S_{\rm F}[\chi, a] -
   \frac{1}{2e_0^2}\sum_{{\bf x},\mu,\nu}\cos\left(F_{\mu\nu}({\bf x})\right).
 \end{equation}
The sites of the three dimensional quadratic lattice are labeled by
$x_{\mu}=\{x_1, x_2, \tau\}$.
Here, $F_{\mu\nu}$ is the usual field-strength tensor
$F_{\mu\nu} = \Delta_{\mu}a_{\nu} -
\Delta_{\nu}a_{\mu}$; the lattice derivative is defined
by $\Delta_{\mu}a_{\nu}({\bf x}) \equiv a_{\nu}({\bf
x}+{\bf \hat{\mu}})-a_{\nu}({\bf x})$.
$S_{\rm F}$ is the lattice action of massless fermions coupled to
the gauge field which reduces in the continuum limit to ${\rm QED}_3$ with
$N_{\rm f}$ flavours of four-component Dirac spinors.
Using staggered fermions, this takes the form
 \begin{equation}
 \label{SFlatt}
   S_{\rm F}[\chi, a] = \frac12\sum_{{\bf x},\mu}\sum_{n=1}^{N_{\rm
   f}/2}\eta_{\mu}({\bf x})\left[\bar{\chi}_n({\bf x})
   e^{ia_{\mu}({\bf x})}\chi_n({\bf x}+{\bf \hat{\mu}}) -
   \bar{\chi}_n({\bf x}+{\bf \hat{\mu}})e^{-ia_{\mu}({\bf x})}\chi_n({\bf x})
   \right]
 \end{equation}
where $\eta_1=1$, $\eta_2=(-1)^{x_1}$ and
$\eta_3=(-1)^{x_1+x_2}$~\cite{Rothe}.

In the case of
continuum ${\rm QED}_3$, the fermion polarisation to one-loop order is
\cite{Pisarski84}
 \begin{equation}
 \label{pol}
   \Pi_{\mu\nu}(p) = \frac{N_{\rm f}}{16}p\left\{\delta_{\mu\nu} -
   \frac{p_{\mu}p_{\nu}}{p^2}\right\}.
 \end{equation}
Incorporating compactness of $a_{\mu}$ in the spirit of  Villain
approximation~\cite{V}, this suggests that we consider a
theory closely related to cQED$_3$
 \begin{equation}
 \label{Seff}
   S[a] = \frac{1}{2}\sum_{{\bf x},\mu,\nu}\left\{
   \left(F_{\mu\nu}({\bf x}) -2\pi n_{\mu\nu}({\bf x}) \right)
   \left(\frac1{2e_0^2}+\frac{N_{\rm f}}{16\left|\Delta\right|}\right)
   \left(F_{\mu\nu} ({\bf x})-2\pi n_{\mu\nu}({\bf x}) \right)\right\}, 
 \end{equation}
where the $n_{\mu\nu}$ are integers.  The action (4)
has the same continuum limit as cQED$_3$ to the leading
order in large $N_{\rm f}$ and may be understood as a compact quadratic approximation
to it. In the remainder of the article we
assume that the original cQED$_3$ and the
theory (4) are in the same universality class.

In the presence of fermions, when $N_{\rm f}\ne0$, the original Maxwell term
proportional to $1/e_0^2$ becomes irrelevant at large distances,
and can be neglected with respect to the second term in Eqn.~(\ref{Seff}).
This action can be then be put into the alternative form (see Appendix~\ref{appmon})
 \begin{equation}
 \label{Zmon}
   Z = \sum
   \exp\left\{-\frac{\pi^2N_{\rm f}}{4}\sum_{a,b}q_aq_bV({\bf x}_a-{\bf x}_b)
   \right\}.
 \end{equation}
This is the partition function for a gas of monopoles of charge $q_{\alpha}
=\pm 1$,
interacting with a potential $V({\bf x})$.  In our case, the potential has the form 
$V({\bf k}) = 1/|k|^3$ in Fourier space,
which is the logarithmic interaction in three dimensions.

The problem now appears to be rather similar to the two dimensional
Coulomb gas, 
where the logarithmic interaction may result in the 
BKT vortex-antivortex binding transition.
The mechanism
of such a transition stems from a simple energy-entropy competition, as both
entropy
and the interaction energy are proportional to the logarithm of linear dimension
of the system: at low enough temperatures, it is energetically favourable
for opposite
vortex charges to form bound pairs, while as temperature is increased, entropy
at some point takes
over, and vortex-antivortex pairs unbind. Although the form of the
partition
function (\ref{Zmon}) resembles that of the two dimensional Coulomb gas,
 it is by no means guaranteed that such a
scenario will still hold in three dimensions. In particular, the effect of
screening of
other dipoles on the potential felt by two widely-separated monopoles, which is
neglected in this na\"{\i}ve energy-entropy argument, can drastically
affect the result.

To systematically address this issue, we first
note that Eq.~(\ref{Zmon}) is equivalent to the
partition function with the anomalous sine-Gordon (ASG) action 
(see Appendix~\ref{appmon})
 \begin{equation}
 \label{ASG}
   S_{\rm ASG}[\phi] = \int{\rm
   d}^3{\bf r}\left\{-\frac{T}{2}\phi\left|\nabla\right|^3\phi -
   2y\cos\phi\right\},
 \end{equation}
where the fictitious temperature is
$T\equiv2/(\pi^2N_{\rm f})$, and $y$ is the fugacity of the monopoles.
The non-analytic gradient term proportional to $|q|^3$ is a consequence of the
coupling of relativistic massless
fermions to the gauge fields.

It is possible to construct
an upper bound for $F_{\rm ASG}$, the free energy associated with
the action (\ref{ASG}), using the
Gibbs-Bogoliubov-Feynman~\cite{Feynman} (GBF)
inequality, which is discussed in the next section. We will argue that 
this self-consistent 
mean-field approximation to the free energy of the
system unfortunately misses the screening effects of the medium,
and consequently incorrectly suggests the BKT transition.  An improved
calculation that incorporates such effects
is then formulated in the following section.

\section{Variational Approach}
\label{sGBF}

The GBF inequality imposes a strict upper bound on the free energy
$F_{\rm ASG}$ through the relation
 \begin{equation}
 \label{GBF}
   F_{\rm ASG} \le F_{\rm var} \equiv F_0 + \left<S_{\rm ASG}-S_0\right>_0  , 
 \end{equation}
where $S_{\rm ASG}$ is defined in Eqn.~(\ref{ASG}) and $S_0$ is a trial action
chosen to approximate $S_{\rm ASG}$; $F_0$ is the free energy associated with
$S_0$ and $\left<...\right>_0$ represents averaging within this ensemble.
The trial action may be chosen to have the Gaussian form  \cite{Giamarchi} 
 \begin{equation}
 \label{SG}
   S_0[\phi] =
   \frac{1}{V}\sum_{\bf {q}}\frac{1}{2}\phi({\bf q})G_0^{-1}({\bf q})\phi(-{\bf
   q}), 
 \end{equation}
so it becomes particularly simple to calculate $F_{\rm var}$:
 \begin{equation}
 \label{Fvar}
   \frac{F_{\rm var}}{V} = -\frac12\int\frac{{\rm d}^3{\bf
   q}}{(2\pi)^3}\ln\left(G_0({\bf q})\right) + \frac{T}{2}\int\frac{
   {\rm d}^3{\bf q}}{(2\pi)^3}|q|^3 G_0({\bf q})
   - 2y~\exp\left\{-\frac{1}{2}\int\frac{{\rm d}^3{\bf
   q}}{(2\pi)^3}G_0({\bf q})\right\}.
 \end{equation}
Minimising $F_{\rm var}$ with respect to $G_0 (q)$ yields the optimal Gaussian
theory that approximates $F_{\rm ASG}$: 
 \begin{equation}
 \label{min}
   \displaystyle\frac{\delta F_{\rm var}}{\delta G_0} = 0
   \Longrightarrow G_0^{-1}({\bf q}) = T|q|^3 + \sigma, 
 \end{equation}
with the `mass' $\sigma$ determined self-consistently through
 \begin{eqnarray}
 \label{sigma}
   \sigma & = & 2y~\exp\left\{-\frac{1}{2}\int\frac{{\rm d}^3{\bf
   q}}{(2\pi)^3}\frac{1}{T|q|^3+\sigma}\right\} \nonumber \\
   & = & 2y\left(1 + \frac{T\Lambda^3}{\sigma}\right)^{-\frac{T_c}{T}}.
 \end{eqnarray}
$\Lambda$ is the ultraviolet cutoff and $T_c \equiv 1/(12\pi^2)$.
Determining the solutions of Eqn.~(\ref{sigma}) amounts to
identifying the roots of the function
 \begin{equation}
 \label{root}
   f(\sigma) = \sigma - 2y\left(1 + \frac{T\Lambda^3}{\sigma}
   \right)^{-\frac{T_c}{T}}.
 \end{equation}
It is evident that $\sigma=0$ is one such root for all values
of $T$.  We next demonstrate that a solution with finite $\sigma$
exists for $T>T_c$.  In the limit of small $\sigma$,
$f(\sigma)$ has the form
 \begin{equation}
   f(\sigma\ll \Lambda^3) = \left\{
   \begin{array}{l}
     \sigma,\quad T<T_c\\
     -\sigma^{T_c/T},\quad T>T_c
   \end{array}\right.
 \end{equation}
while for large $\sigma$
 \begin{equation}
   f(\sigma\gg \Lambda^3) = \sigma, \quad \forall~T.
 \end{equation}
For $T>T_c$, $f(\sigma)$ changes sign and thus has a root with $\sigma>0$,
while only the $\sigma=0$ solution exists for $T<T_c$~\cite{Diehl97}.

The stability of the $\sigma=0$ solution for $T>T_c$ can be determined
from the variational free energy~(\ref{Fvar}) with the
solution~(\ref{min}) for $G_0^{-1}$.  Evaluating the free energy we get
 \begin{equation}
   \frac{F_{\rm var}(\sigma)}{V} = T_c\Lambda^3\ln\left(\sigma+T\Lambda^3\right)
   - 2y\left(1+\frac{T\Lambda^3}{\sigma}\right)^{-T_c/T}.
 \end{equation}
Then
 \begin{equation}
   \frac{1}{V}\left(F_{\rm var}(\sigma)-F_{\rm var}(0)\right) =
   \frac{\sigma(T_c-T)}{T} + O(\sigma^2),
 \end{equation}
so that for $T>T_c$ any solution with $\sigma>0$ is of lower free energy
than with $\sigma=0$.  That is, the stable solution at $T>T_c$ has
finite $\sigma$.

To understand the physical meaning of the non-trivial solution it is useful to
calculate the monopole density from the variational free
energy~(\ref{Fvar}):
 \begin{eqnarray}
   \rho_{M} &=& -\frac1{V}\frac{\partial F_{\rm var}}{\partial
   \mu}\nonumber \\
   &=& -\frac{y}{V}\frac{\partial F_{\rm var}}{\partial y}\nonumber \\
   &=& \sigma, 
 \end{eqnarray}
where we have used the definition of fugacity $y \equiv \exp\{\mu\}$.
We see that $\sigma$ is exactly the monopole density $\rho_M$, so that
$\sigma\neq 0$ may be identified with the plasma phase of
free monopoles, while $\sigma=0$ indicates the  dipole phase. 
The simple variational calculation would therefore suggest
that monopoles undergo a binding-unbinding transition at $T=T_c$
(i.~e. at $N=N_c = 24$) in exact analogy with the equivalent
calculation one can perform for the standard BKT transition.
The value of $T_c$ also agrees with the simple energy-entropy argument
that can be constructed for an isolated vortex \cite{Igor03}. 

An obvious objection to this simple calculation is that minimisation
of the variational free energy (9) by construction
cannot yield any momentum dependence
of the self-energy, but can only determine its constant part, the `mass'
$\sigma$. The renormalisation group \cite{Igor03} treatment of the  ASG
theory suffers from the same problem to the lowest order in fugacity,
and would likewise na\"{\i}vely suggest the BKT transition. The same holds for
the direct perturbative evaluation of the self-energy in the
ASG. However, it is easy to check that the self-energy does become
momentum dependent to the {\it second} order in fugacity, 
with the leading analytic term $\sim q^2$ at low momenta.
This is just what one would expect
based on the simple electrostatic analysis of the problem \cite{Igor03},
where this term translates into
the Coulombic interaction in real space (when $y=0$).
The presence of such a term would, however, drastically alter 
our present considerations. Indeed, if we
add by hand the term $ Q q^2$ with $Q\neq 0$
in the denominator of the integrand
in the self-consistent equation~(\ref{sigma}), we find
 \begin{equation}
   f(\sigma) = \left\{
   \begin{array}{l}
     -2y\left(1+\frac{\Lambda T}{Q}\right)^{-3T_c/T},\quad\sigma\ll
     \Lambda^3\\
     \sigma,\quad\sigma\gg\Lambda^3 
   \end{array}
   \right.
 \end{equation}
for all $T$.  Hence, the non-trivial solution would exist for all
temperatures,
exactly as in Polyakov's original treatment of the pure gauge
theory. This is natural since $Q\neq 0$ means that the original logarithmic
interaction between monopoles is, even without free monopoles and only with
a finite density of dipoles,
screened into the Coulomb interaction for which the standard argument for
the confined phase readily applies. 

In the next section we propose a modified self-consistent calculation 
which provides a systematic perturbative approximation
to the free energy and which reduces to 
the GBF method to the lowest order.  As we will see in
Section~\ref{results}, such an approach has
the advantage of including the screening effects in a
self-consistent way, therefore overcoming the limitations of the
purely variational theory  discussed in this section.

\section{Self-consistent perturbative approach}
\label{SCPA}

There are many ways in which one may generalise the variational method of
the previous
section. For instance, one may add a second-order term $-\frac12\langle
{(S_{\rm ASG}-S_0)^2\rangle}_0+\frac12\langle S_{\rm
ASG}-{S_0\rangle}_0^2$ to $F_{\rm var}$
and extremise the new energy functional~\cite{Feynman}.
Such a second-order extension, however, 
has little variational justification. For a more systematic generalisation,
we go back
to the GBF inequality~(\ref{GBF}) and exchange $S_{\rm ASG}$ with $S_0$ to find
 \begin{equation}
 \label{GBF<}
   F_{<} \equiv F_0+\langle S_{\rm ASG}-S_0 \rangle \le F_{\rm ASG}.
 \end{equation}
Extremising $F_<$ with respect to a quadratic action $S_0$ yields
 \begin{eqnarray}
 \label{SCP}
   \langle\phi(-{\bf q}){\phi({\bf q})\rangle}_0=\langle\phi(-{\bf
   q})\phi({\bf
   q})\rangle,
 \end{eqnarray}
which is nothing but the equation for the {\it exact}
 propagator in the ASG theory. The right hand side (RHS) of the equation,
 on the other hand may be rewritten as
 \begin{eqnarray}
 \label{fullprop}
   \langle\phi(-{\bf q})\phi({\bf q})\rangle \equiv
   \frac{\langle\phi(-{\bf q})\phi({\bf q}) {e^{-\Delta S}\rangle}_0}{\langle
   {e^{-\Delta S}\rangle}_0}, 
 \end{eqnarray}
with $\Delta S\equiv S_{\rm ASG}-S_0$. Eqn. (18) in this form may
be understood as 
a self-consistent equation for the action $S_0$, which we may
attempt to solve by expanding the RHS in powers of $\Delta S$, for example.
To the first order in $\Delta S$ this becomes
 \begin{eqnarray}
   \langle\phi(-{\bf q})\phi({\bf q})\Delta {S\rangle}_0 - \langle\phi(-{\bf
   q}){\phi({\bf q})\rangle}_0\langle\Delta {S\rangle}_0 = 0, 
 \end{eqnarray}
which is precisely the relation one would obtain from extremising $F_{\rm
var}$ with respect to $S_0$.  That is, the first order approximation to
Eqn.~(\ref{SCP}) reproduces the GBF result from the previous section.

Eqn.~(\ref{SCP}) forms the basis of our modified variational approximation to
$F_{\rm ASG}$.
To the first order in $\Delta S$ it reduces to the GBF equation of the
previous section, and when solved self-consistently to all orders gives the best
variational lower bound to the free energy, provided by $F_<$ in~(\ref{GBF<}). In
addition, consider the expansion of Eqn.~(\ref{SCP}) to order
$(\Delta S)^n$. One can show (see Appendix~\ref{Fvarnapp}) that the 
resulting expression is the same as the one that would arise from extremising the function
 \begin{equation}
 \label{Fn}
   F_{\rm var}^{(n)} \equiv \frac{F^{(1)}+F^{(2)}+
   \cdots+F^{(n)}}{n}.
 \end{equation}
Here $F^{(n)}$  stands for the expansion of the true free
energy of the system, $F_{\rm ASG}$, in powers
of $\Delta S$, truncated at $(\Delta S)^n$.
Similarly denoting by $F^{(n)}_<$ the truncated
expansion of $F_<$ in Eqn.~(\ref{GBF<}), it is not difficult
(see Appendix~\ref{F<napp}) to show
 \begin{equation}
 \label{F<n}
  F_{\rm var}^{(n)} = F^{(n)}+\frac{F^{(n)}-F_<^{(n)}}{n}.
 \end{equation}
It is then clear that the sequence
$\{F^{(n)}_{\rm var}\}$ converges to  $F_{\rm ASG}$ for any $S_0$.
Therefore, the
$S_0$ determined self-consistently from Eqn.~(\ref{SCP}) also yields the
variational sequence that best approximates $F_{\rm ASG}$ within the family
$\{F^{(n)}_{\rm var}[S_0]\}$.

To the second order Eqn.~(\ref{SCP}) reads
 \begin{equation}
 \label{2ndord}
   \langle\phi(-{\bf q})\phi({\bf q})\Delta {S\rangle}_0^{\rm c}
   -\frac12\langle\phi(-{\bf q}){\phi({\bf q})(\Delta S)^2\rangle}_0^{\rm
   c}=0,
 \end{equation}
where both terms are connected averages given by:
 \begin{eqnarray}
 \label{dSc}
   \langle\phi(-{\bf q})\phi({\bf q})\Delta {S\rangle}_0^{\rm c} &\equiv&
   \langle\phi(-{\bf q})\phi({\bf q})\Delta {S\rangle}_0
   -\langle\phi(-{\bf q}){\phi({\bf q})\rangle}_0\langle\Delta
   {S\rangle}_0,\\
   \langle\phi(-{\bf q}){\phi({\bf q})(\Delta S)^2\rangle}_0^{\rm c} &\equiv&
   \langle\phi(-{\bf q}){\phi({\bf q})(\Delta S)^2\rangle}_0
   -\langle\phi(-{\bf q}){\phi({\bf q})\rangle}_0\langle{(\Delta
   S)^2\rangle}_0
   \nonumber\\
 \label{dS2c}
   && -2\langle\phi(-{\bf q})\phi({\bf q})\Delta {S\rangle}_0\langle\Delta
   {S\rangle}_0
   +2\langle\phi(-{\bf q}){\phi({\bf q})\rangle}_0\langle\Delta
   {S\rangle}_0^2.
 \end{eqnarray}
We discuss the results of the second-order self-consistent
Eqn.~(\ref{2ndord})
for the ASG model~(\ref{ASG}) in the next section.  In particular, we will 
show that the density of free monopoles is finite at all $T>0$, and 
that charge should consequently be  permanently confined in ${\rm cQED}_3$.

\section{Confining Solution for $T>0$}
\label{results}

From the definitions of $S_{\rm ASG}$ and $S_0$ (Eqns.~\ref{ASG} and~\ref{SG})
it is straightforward to calculate the connected averages of
Eqns.~(\ref{dSc}) and~(\ref{dS2c}).  Our second order
equation~(\ref{2ndord}) then yields the quadratic equation for
$G_0^{-1}({\bf q})$
 \begin{eqnarray}
 \label{quad}
   \left[G_0^{-1}({\bf q})\right]^2 - A[{\bf q},G_0]G_0^{-1}({\bf q})
   +B[{\bf k},G_0]=0,
 \end{eqnarray}
where
 \begin{eqnarray}
 \label{A}
   A[{\bf q}, G_0]&=&\frac32T|q|^3
   +3a+ab-2a^2\left(c+\sum_{n=0}^{\infty}(-1)^nd_nq^{2n}\right)\\
 \label{B}
   B[{\bf q}, G_0] &=& \frac12T^2q^6+2aT|q|^3.
 \end{eqnarray}
In Eqns.~(\ref{A},~\ref{B}), we have defined
 \begin{eqnarray}
 \label{a}
   a &=& ye^{-\frac12D_0(0)},\\
 \label{b}
   b &=& \int\frac{{\rm d}^3{\bf k}}{(2\pi)^3}\left(\frac{T}2|k|^3-
   \frac12G_0^{-1}({\bf
   k})\right)\left[G_0({\bf k})\right]^2,\\
 \label{c}
   c &=& \int{\rm d}^3{\bf R}\left[1 - \cosh D_0({\bf
   R})\right],\\
 \label{d}
   d_n &=& \int{\rm d}^3{\bf
   R}\frac{(R\cos\theta)^{2n}}{(2n)!}\sinh D_0({\bf R}),
 \end{eqnarray}
and the real-space propagator is $D_0({\bf R}) = \int{\rm d}^3{\bf k}
/(2\pi)^3G_0({\bf k})e^{i{\bf k}\cdot{\bf R}}$.

We can solve the quadratic of Eqn.~(\ref{quad}) and expand in powers of
$|q|^3/A_0$ to yield the result for $G_0^{-1}({\bf q})$:
 \begin{equation}
 \label{G0inv}
   G_0^{-1}({\bf q}) = m + Q(m)q^2 + \widetilde{T}(m)|q|^3 + \cdots
 \end{equation}
where the coefficients are defined as
 \begin{eqnarray}
 \label{ml}
   m &=& \frac12\left\{A_0 \pm \left|A_0\right|\right\},\\
 \label{Ql}
   Q(m) &=& a^2d_1\left(1 \pm \frac{\left|A_0\right|}{A_0}\right),\\
 \label{Tl}
   \widetilde{T}(m) &=& \frac34T \pm
   \frac{\left|A_0\right|}{A_0}\left(\frac34T-\frac{2aT}{A_0}\right);
 \end{eqnarray}
and with $A_0 \equiv A[q=0,G_0]$.
For these equations, we should choose the solution corresponding to the
upper sign in Eqns.~(\ref{ml}~--~\ref{Tl}) to ensure that $m\ge 0$.
In what follows, we neglect terms higher order in $q$ than $q^3$ as they
should be irrelevant at low momenta.

As announced, the second order result includes additional renormalisation
of the bare terms as well as the generation of new momentum dependent terms.
Most importantly, the leading term proportional to $q^2$ has now
appeared.

In the  analysis in Section~\ref{sGBF} we found that the bound phase of
monopoles corresponded to low $T$. In what follows we will restrict
ourselves to low temperatures by assuming 
$T\Lambda\ll Q$ and show that monopoles are unbound even
for arbitrarily small temperatures. By continuity this would imply that 
they are free at all temperatures.

 Let us start by examining $a$:
 \begin{eqnarray}
 \label{acalc}
   a &=& y~\exp\left\{-\frac12D_0(0)\right\}\nonumber\\
   &\approx& y~\exp\left\{-\frac12\int\frac{{\rm d}^3{\bf k}}{(2\pi)^3}
   ~\frac1{\left(Q(m)k^2+m\right)}\right\}\nonumber\\
   &=& y~\exp\left\{-\frac1{4\pi^2Q(m)}\left(\Lambda -
   \sqrt{\frac{m}{Q(m)}}\arctan\left(\Lambda\sqrt{\frac{Q(m)}{m}}. 
   \right)\right)\right\},
 \end{eqnarray}
When $m\rightarrow0$, we will assume $m/Q(m)\rightarrow 0$, and 
justify this assumption {\it a posteriori}.
The coefficient $a$ now takes the form
 \begin{equation}
 \label{asmallm}
   a = y~\exp\left\{-\frac{\Lambda}{4\pi^2Q(m)}\right\}
   ,\quad m\ll\Lambda^3  +O(T).
 \end{equation}

Next, we examine the equation for $b$
 \begin{eqnarray}
 \label{bcalc}
   b &=& \int\frac{{\rm d}^3{\bf k}}{(2\pi)^3}\left(\frac{T}2|k|^3
   -\frac12G_0^{-1}({\bf
   k})\right)\left[G_0({\bf k})\right]^2\nonumber\\
   &=& -\frac12D_0(0) + O(T).
 \end{eqnarray}
From this we find 
 \begin{equation}
   b = -\frac{\Lambda}{4\pi^2Q(m)},\quad m\ll\Lambda^3 +O(T).
 \end{equation}

Next, as the terms $c$ and $d_0$ always appear together, we consider the
combination
 \begin{eqnarray}
 \label{cd0}
   (c+d_0) &\approx& \int{\rm d}^3{\bf R}\left(1-\exp\left\{-\int\frac{{\rm
   d}^3{\bf k}}{(2\pi)^3}~\frac{e^{i{\bf k}\cdot{\bf
   R}}}{\left(Q(m)k^2 + m\right)}
   \right\}\right)+O(T) \nonumber\\
   &=& \int_0^{\infty}{\rm d}R
   ~\frac{R{\rm e}^{-\sqrt{\frac{m}{Q(m)}}R}}{Q (m)} +O(T) .
 \end{eqnarray}
Evaluating this yields
 \begin{equation}
   (c+d_0) = m^{-1},\quad m\ll\Lambda^3 +O(T).
 \end{equation}

Similar analysis applies to the coefficient $d_1$:
 \begin{eqnarray}
 \label{d1}
   d_1 = \frac16\int_0^{\infty}{\rm d}R~
   \frac{R^3{\rm e}^{-\sqrt{\frac{m}{Q(m)}}R}}{Q(m)} +O(T), 
 \end{eqnarray}
which gives
 \begin{equation}
   d_1 = \frac{Q (m)}{m^2},\quad m\ll\Lambda^3 +O(T).
 \end{equation}

Evaluating the Eqn.~(\ref{Ql}) for $Q$ then we find
 \begin{eqnarray}
   Q &=& 2a^2d_1\nonumber \\
   &=& 2y^2\exp\left\{-\frac{\Lambda}{2\pi^2Q}\right\}\frac{Q}{m^2} +O(T) .
 \end{eqnarray}
Solving this for $Q\neq 0 $ yields
 \begin{equation}
   Q = \frac{\Lambda}{4\pi^2}\left(\ln \frac{\sqrt2y}{m}\right)^{-1} +O(T),
 \end{equation}
and we see that $m/Q(m)$ indeed
approaches zero as $m \rightarrow 0$, thus
justifying our earlier assumption.  Substituting this solution for $Q(m)$
into our mass equation (\ref{ml}) gives
 \begin{eqnarray}
   m &=& A_0\nonumber\\
   &\approx& \frac{m}{\sqrt2}\left[3-\sqrt2-\ln\frac{\sqrt2y}{m}\right], 
 \end{eqnarray}
which can finally be solved for $m\neq 0 $ to give the finite mass solution
 \begin{equation}
   m^*=\sqrt2 {\rm e}^{2\sqrt2-3} y . 
 \end{equation}
The corresponding finite value of $Q$ is 
 \begin{equation}
   Q^*=\frac{\Lambda}{2\pi^2(3-2\sqrt2)}.
 \end{equation}
Note that $m^*$ is proportional to $y$ so that small fugacity translates to
small $m^*$, in accord with our assumption that  $m\ll\Lambda^3$.

To show that monopoles are free when $m\neq 0$, we calculate the monopole
density as in Section~\ref{sGBF}.  From Eqn.~(\ref{Fn}) we see that the free
energy associated with our second order equation~(\ref{2ndord}) is
 \begin{equation}
 \label{Fvar2}
   F_{\rm var}^{(2)} = F_0 + \langle\Delta {S\rangle}_0-\frac14\langle{(\Delta
   S)^2\rangle}_0 + \frac14\langle\Delta {S\rangle}_0^2.
 \end{equation}
From this, the monopole density can be calculated.
 \begin{eqnarray}
   \rho_M^{(2)} &=& -\frac1{V}\frac{\partial F_{\rm var}^{(2)}}{\partial
   \mu}\\ \nonumber
   &=& 2a+ab-2a^2c.
 \end{eqnarray}
For $m=0$, the monopole density vanishes, while for our finite $m$ solution
 \begin{eqnarray}
   \rho_M^{(2)} &=&
   \frac{m^*}{\sqrt{2}}\left(2\sqrt{2}-1+
   \frac1{16\pi}\sqrt{\frac{2m^*}{(Q^*)^3}}
   \right)\nonumber\\
   &>&0.
 \end{eqnarray}
From the free energy~(\ref{Fvar2}), it is also possible to show that the
finite $m$ solution is the stable solution for all temperatures.  In fact,
the free energy diverges as $\log(1/m)$ as $m$ approaches zero, but has a
finite value for finite $m$.  It is then the free phase of monopoles which
is favoured at all temperatures.

Thus we have demonstrated that for arbitrarily low $T$ a finite mass solution
always exists for the self-consistent equations~(\ref{ml}~--~\ref{Tl}).
This implies that monopoles are always free at low temperatures,
or, in terms of the
original lattice model~(\ref{Slatt}), that the electric charge is presumably
confined for any number of fermion flavours.

\section{Conclusion}
\label{conc}

We have studied ${\rm cQED}_3$ where massless relativistic
fermions coupled to the compact
gauge field result in a logarithmic interaction between magnetic monopoles.
One may suspect that this could lead to a BKT-like transition where free
monopoles bind into monopole-antimonopole pairs
at low enough effective temperatures.  Although the simplest
mean-field approximation would predict such a transition, we argued that
by design this treatment misses the screening effects, argued to be
crucial in this problem. To address
this issue we developed a combined variational-perturbative
approach which allowed us to include screening self-consistently.
The modified theory then leads to the plasma phase of free monopoles
as being stable at all temperatures, in agreement with the renormalisation
group treatment of the problem~\cite{Igor03}.

  cQED$_3$ has been studied numerically in~\cite{Fiebig} 
and~\cite{Azcoiti}.  Our calculation appears to be in agreement with
the numerical results of~\cite{Azcoiti}, where only a single phase was
observed. We hope that this and previous work on cQED$_3$ will motivate
renewed efforts in this direction, using bigger system sizes that
have recently become available~\cite{Hands}.

\section{Acknowledgment}

This research has been supported by the NSERC of Canada and the Research
Corporation.

\appendix
\section{Monopole gas and the sine-Gordon action}
\label{appmon}

In this appendix, we fill in the
details in going from the action~(\ref{Seff}) to the Coulomb gas partition
function~(\ref{Zmon}) and the sine-Gordon action~(\ref{ASG}). For this
purpose, let us
write~(\ref{Seff}) on the lattice in a more general form as
 \begin{equation}
   S[a,n] = \frac12\sum_{{\bf x},{\bf y}} (F_{\mu\nu}
   ({\bf x})-2\pi n_{\mu\nu}({\bf x})) \;
   u({\bf x},{\bf y})\;(F_{\mu\nu}({\bf y})-2\pi n_{\mu\nu}({\bf y})),
 \end{equation}
where $u^{-1}({\bf x},{\bf y})=(16/N_{\rm f})|\Delta_{\bf x}|\delta_{{\bf x},{\bf y}}$ and
$\Delta_{{\bf x},\mu} f \equiv f({\bf x}+\hat\mu)-f({\bf x})$ denotes the
lattice derivative.
Introducing an antisymmetric Hubbard-Stratonovich field $M_{\mu\nu}$ we
find
 \begin{eqnarray}
   S &\to& \frac12\sum_{{\bf x},{\bf y}}M_{\mu\nu}({\bf x})\;u^{-1}({\bf
   x},{\bf
   y})\;M_{\mu\nu}({\bf y}) + i\sum_{\bf x}M_{\mu\nu}({\bf
   x})(F_{\mu\nu}({\bf x})-2\pi n_{\mu\nu}({\bf x})) \nonumber \\
   &=&   \frac14\sum_{{\bf x},{\bf y}}b_\mu({\bf x})u^{-1}({\bf x},{\bf
   y})b_\mu({\bf y})
   + i \sum_{\bf x}(\nabla\times{\bf a}-2\pi {\bf n})(\bf
   x)\cdot{\bf b}(\bf x),
 \end{eqnarray}
with $b_\mu\equiv\epsilon_{\mu\nu\lambda}M_{\nu\lambda}$ and
$n_\mu\equiv\epsilon_{\mu\nu\lambda}n_{\nu\lambda}$. Integrating over the
gauge field constrains the
${\bf b}$-field to be curl-free, so we can take it to be a gradient on the
lattice ${\bf b}=\Delta\varphi$.
Performing the lattice version of integration by parts and integrating over
$\varphi$ yields
 \begin{eqnarray}
   S &\to& \frac14\sum_{{\bf x}, {\bf y}}\Delta_\mu\varphi({\bf
   x})\;u^{-1}({\bf x},{\bf y})\;\Delta_\mu\varphi({\bf y})
   + 2\pi i \sum_{\bf x}\Delta\cdot{\bf n}({\bf x})\;\varphi({\bf x})\\
   &\to& \frac12\sum_{{\bf x}, {\bf y}}\rho({\bf x})\;v({\bf x},{\bf
   y})\;\rho({\bf y}),
 \end{eqnarray}
where $v^{-1}({\bf x}, {\bf y}) = -(1/8\pi^2)\Delta_{{\bf x},\mu}
u^{-1}({\bf x}, {\bf y})\Delta_{{\bf y},\mu}$
is the (inverse of the) potential and $\rho=\Delta_\mu n_\mu$ is
the density of magnetic monopoles. Using the expression for $u({\bf x},
{\bf y})$ we
find, in the continuum limit,
 \begin{eqnarray}
   v({\bf x},{\bf y}) &=& \frac{\pi^2N_{\rm f}}2\int
   \frac{{\rm d}^3{\bf k}}{(2\pi)^3}
   \frac{e^{i{\bf k}\cdot({\bf x}-{\bf y})}}{k^3}
   \nonumber\\
   &\equiv& \frac{\pi^2N_{\rm f}}2 V({\bf x}-{\bf y}).
 \end{eqnarray}
Thus, for a system of $N$ monopoles with a density
$\rho({\bf x})=\sum_{a=1}^{N}q_a\delta({\bf x}-{\bf x}_a)$, we obtain
 \begin{equation}
 \label{Smon}
   S_{\rm mon}= \frac{\pi^2 N_{\rm f}}{4}\sum_{a, b}q_a q_b V({\bf
   x}_a-{\bf x}_b),
 \end{equation}
as in Eqn.~(\ref{Zmon}).

We now proceed to show that~(\ref{Smon}) with $q_a=\pm 1$ is equivalent to
the sine-Gordon action~(\ref{ASG}). Higher charges are irrelevant for large
enough $N_{\rm f}$.
To this end, let us introduce the bare action
 \begin{equation}
   S_{\rm b} = \frac12\int {\rm d}^3{\bf x}\;{\rm d}^3{\bf y}\; \phi({\bf
   x})v^{-1}({\bf x}-{\bf y})\phi({\bf y}),
 \end{equation}
so that $v({\bf x}-{\bf y})=\langle\phi({\bf x})\phi({\bf y})\rangle_{\rm
b}$. Showing the
fugacity by $y$, we may write the monopole partition function in the
grand-canonical ensemble as
 \begin{eqnarray}
   Z_{\rm mon} &=& \sum_{N}\frac{y^N}{N!}\int\prod_i^N{\rm d}^3{\bf
   x}_i\sum_{\{q_a=\pm1\}}
   \exp\left\{-\frac12\sum_{a,b}q_a q_b\left<\phi({\bf
   x}_a)\phi({\bf x}_b)\right>_{\rm b}\right\}
   \nonumber\\
   &=& \sum_{N}\frac{y^N}{N!}\int\prod_i^N{\rm d}^3{\bf
   x}_i\sum_{\{q_a=\pm1\}}
   \left\langle \exp\left[i\sum_a q_a\phi({\bf
   x}_a)\right]\right\rangle_{\rm b} \nonumber\\
   &=& \left< \exp \left\{ 2y\int{\rm d}^3{\bf x}\; \cos\phi({\bf
   x}) \right\} \right>_{\rm b}
   \nonumber\\
   &\equiv& Z_{\rm b}^{-1}Z_{ASG},
 \end{eqnarray}
where $Z_{\rm b}$ is independent of $\phi$ and
 \begin{equation}
   Z_{ASG} = \int\mathcal{D}\phi\;
   \exp \left\{ -\frac12 \int {\rm d}^3{\bf x}\;{\rm d}^3{\bf y}\;
   \phi({\bf x})v^{-1}({\bf x}-{\bf y})\phi({\bf y})
   + 2y\int{\rm d}^3{\bf x} \cos\phi({\bf x}) \right\}.
 \end{equation}
Inserting the definition of $v^{-1}({\bf x}-{\bf y})$ into this last
expression we immediately arrive at the anomalous sine-Gordon action
(\ref{ASG}).

\section{}
\subsection{$F_<^{(n)}$}
\label{F<napp}

Here we will show that $F_<^{(n)}$ indeed satisfies Eqn.~(\ref{F<n}). 
To this end, first let us define 
$\Delta F^{(n)} \equiv F^{(n+1)}-F^{(n)}$. Then we may equivalently show that
 \begin{equation}
 \label{F<nmod}
   F_<^{(n)}=F_0+\Delta F^{(1)}+2\Delta F^{(2)}+\cdots+n\Delta F^{(n)}.
 \end{equation}
Let us also denote the path integral over the field $\phi({\bf q})$ by
${\rm Tr}$ and define, for a real variable~$t$,
 \begin{equation}
 \label{Ft}
   \mathcal F(t) \equiv -\ln{\rm Tr}\left(e^{-S_0}e^{-t\Delta S}\right).
 \end{equation}
Then, $\mathcal F(1)=-\ln{\rm Tr}\;\exp(-S)=F_{\rm ASG}$ and
 \begin{equation}
   \left.\frac{\d\mathcal F(t)}{\d t}\right\vert_{t=1} = 
   \frac{{\rm Tr}\left(\Delta S\;e^{-S}\right)}{{\rm Tr}\left(e^{-S}\right)} = 
   \langle\Delta S\rangle.
 \end{equation}
On the other hand, we may expand the RHS of Eqn.~(\ref{Ft}) in powers of
$\Delta S$ as
 \begin{equation}
   \mathcal F(t) = F_0 + \sum_{i=1}^\infty \Delta \mathcal F^{(i)}(t)
 \end{equation}
where $\Delta \mathcal F^{(i)}(t)=t^i\Delta \mathcal
F^{(i)}(1)=t^i\Delta F^{(i)}$. Thus
 \begin{equation}
 \label{F'}
   \langle\Delta S\rangle = \left.\frac{\d\mathcal F(t)}{\d t}
   \right\vert_{t=1} =
   \sum_{i=1}^\infty i\;\Delta F^{(i)}.
 \end{equation}
Upon insertion of Eqn.~(\ref{F'}) into the definition of $F_<$ in 
Eqn.~(\ref{GBF<}) and truncating the expansion at $i=n$ we
find~(\ref{F<nmod}). 

\subsection{$F_{\rm var}^{(n)}$}
\label{Fvarnapp}

In this appendix, we will give the proof for our claim that the extremum
of $F_{\rm var}^{(n)}$ as defined in (\ref{Fn}) is given by the expansion of
Eqn.~(\ref{SCP})
to order $(\Delta S)^n$, i.~e.
 \begin{equation}
 \label{claim}
   \frac{\delta F_{\rm var}^{(n)}}{\delta G_0({\bf k})}
   =0 \;\Longleftrightarrow\;
   \langle\phi(-{\bf k}){\phi({\bf k})\rangle}_0 = \langle\phi(-{\bf k})
   \phi({\bf k})\rangle^{(n)}.
 \end{equation}
The calculations are, for general $n$, cumbersome and not very instructive
so we will first present the case for $n=2$ which is also the one with which
we are concerned in Section~\ref{results}.
 
Setting $n=2$, we see that Eqn.~(\ref{2ndord}) is readily found by an expansion
of the RHS of Eqn.~(\ref{fullprop}).  To show that the same result arises from
extremising $F_{\rm var}^{(n)}$, it is first useful to establish
 \begin{eqnarray}
 \label{dF_0dG_0}
   \frac{\delta F_0}{\delta G_0({\bf q})} 
   &=& \left<\frac{\delta S_0}{\delta G_0({\bf q})}\right>_0 
   = \frac{-1}{2(2\pi)^3[G_0({\bf q})]^2}
   {\langle\phi(-{\bf q})\phi({\bf q})\rangle}_0, \\ 
   \frac{\delta {\langle g\rangle}_0}{\delta G_0({\bf q})}
   &=& \frac{\delta F_0}{\delta G_0({\bf q})}{\langle g\rangle}_0 +
   \left<\frac{\delta g}{\delta G_0({\bf q})}-g\frac{\delta S_0}
   {\delta G_0({\bf q})}\right>_0,
 \end{eqnarray}
where $g=g(S_0)$ is an arbitrary function of $S_0$. Thus choosing appropriate
forms of $g$
for $F^{(1)} = F_0 + {\langle\Delta S\rangle}_0$ and 
$F^{(2)} = F^{(1)}-\frac12 \langle(\Delta {S)^2\rangle}_0
+\frac12\langle\Delta {S\rangle}_0^2$
we find
 \begin{eqnarray}
 \label{dF1}
   \frac{\delta F^{(1)}}{\delta G_0({\bf q})} 
   &=& \frac{\delta F_0}{\delta G_0({\bf q})}{\langle \Delta S\rangle}_0 
   - \left<\Delta S\frac{\delta S_0}{\delta G_0({\bf q})}\right>_0,\\
   \frac{\delta F^{(2)}}{\delta G_0({\bf q})}
   &=& \frac{\delta F_0}{\delta G_0({\bf q})}
   \left[-\frac12\langle\Delta {S^2\rangle}_0+\langle\Delta
   {S\rangle}_0^2\right]
   \nonumber\\
 \label{dF2}
   &+&  \frac12\left<\frac{\delta S_0}{\delta G_0({\bf q})}(\Delta S)^2\right>_0
   -\left<\frac{\delta S_0}{\delta G_0({\bf q})}\Delta S\right>_0
   \langle\Delta {S\rangle}_0.
 \end{eqnarray}
Inserting Eqn.~(\ref{dF_0dG_0}) into Eqns.~(\ref{dF1},~\ref{dF2}) and
adding them we find that the restriction
$\delta F^{(2)}_{\rm var}/\delta G_0({\bf q})=0$ leads to the same equation as
Eqn.~(\ref{2ndord}).

The proof for arbitrary $n$ goes along essentially the same steps as above.
Various 
truncated expansions we have defined can be read off the Taylor expansion
identity
 \begin{equation}
 \label{Taylorln}
   -\ln{\rm Tr}\left\{ e^{-S_{\rm b}-V}\right\} = F_{\rm b} + \sum_{i=1}^\infty
   \sum_{l=1}^i\frac{(-1)^{i+l}}l {\sum_{\{k_\alpha\}}}'
   \frac{\langle{V^{k_1}\rangle}_{\rm b}\cdots\langle{V^{k_l}\rangle}_{\rm b}}
   {k_1!\cdots k_l!},
 \end{equation}
by setting $i$ to the desired order. 
In~(\ref{Taylorln}) $S_{\rm b}$ and $V$ give an arbitrary splitting of the
action into a bare and potential part respectively and
 \begin{equation}
   {\sum_{\{k_\alpha\}}}' \equiv \sum_{k_1=1}^{i}\cdots\sum_{k_l=1}^{i}
   \delta_{k_1+\cdots+k_l,i}.\nonumber
 \end{equation}
Notice that in Eqns.~(\ref{dF1}) and~(\ref{dF2}) all the terms are to the
same order of $\Delta S$, which is also the largest in the corresponding
expansion of the free energy.
By choosing $S_{\rm b}=S_0$ and $V=\Delta S$ and setting $i=n$
in~(\ref{Taylorln})
one can see,  after some lengthy algebra, that the same is true for arbitrary
$n$:
 \begin{eqnarray}
 \label{Fnexpan}
   \frac{\delta F^{(n)}}{\delta G_0({\bf q})}
   &=& \sum_{l=1}^n(-1)^{n+l}{\sum_{\{k_\alpha\}}}'
   \frac{\langle(\Delta {S)^{k_1}\rangle}_0\cdots 
   \langle(\Delta {S)^{k_l}\rangle}_0}{k_1!\cdots k_l!}
   \frac{\delta F_0}{\delta G_0({\bf q})} \nonumber\\
   &-& \sum_{l=1}^n(-1)^{n+l}{\sum_{\{k_\alpha\}}}'
   \frac{\left<\frac{\delta S_0}{\delta G_0({\bf q})}(\Delta S)^{k_1}\right>_0
   \langle(\Delta {S)^{k_2}\rangle}_0\cdots 
   \langle(\Delta {S)^{k_l}\rangle}_0}{k_1!k_2!\cdots k_l!}.
 \end{eqnarray}

Let us now define, for a real variable $t$,
 \begin{equation}
   \mathcal G({\bf k},t) \equiv -\ln{\rm Tr}\left\{e^{-S_0-\Delta S
   -t\phi(-{\bf k})\phi({\bf k})}\right\},
 \end{equation}
so that
$\partial \mathcal G({\bf k}, t)/\partial t|_{t=0}=\langle\phi(-{\bf k})
\phi({\bf k})\rangle$.
Then, taking $S_{\rm b}=S_0$ and $V=\Delta S+t\phi(-{\bf k})\phi({\bf k})$ in 
Eqn.~(\ref{Taylorln})
to compute this derivative,
it can be shown through additional tedious but straightforward algebra that
 \begin{equation}
   \langle\phi(-{\bf q})\phi({\bf q})\rangle^{(n)}-
   \langle\phi(-{\bf q}){\phi({\bf q})\rangle}_0 = 
   -2n(2\pi)^3[G_0({\bf q})]^2\frac{\delta F_{\rm var}^{(n)}}
   {\delta G_0({\bf q})},
 \end{equation}
where we have also made use of Eqn.~(\ref{Fnexpan}).  Thus, the requirement
that $F_{\rm var}^{(n)}$
be an extremum implies Eqn.~(\ref{SCP}) truncated at $n$th order, and
{\it vice versa}, proving our
claim~(\ref{claim}).

\end{document}